\documentclass[a4paper,10pt]{article}
\usepackage{pos}

\def\ol{\overline}

\def\be{\begin{eqnarray}}
\def\en{\end{eqnarray}}
\def\non{\nonumber}
\def\CP{{\it CP}~}

\def\la{\langle}
\def\ra{\rangle}

\def\PE{{P\!E}}
\def\PA{{P\!A}}
\def\gsim{ {\ \lower-1.2pt\vbox{\hbox{\rlap{$>$}\lower5pt\vbox{\hbox{$\sim$}
}}}\ } }

\title{Direct {\it CP} violation in charmed meson decays within the standard model}

\author*[a]{Hai-Yang Cheng}
\author[b]{Cheng-Wei Chiang}

\affiliation[a]{Institute of Physics, Academia Sinica,\\
  Taipei, Taiwan 11529, ROC}

\affiliation[b]{Department of Physics, National Taiwan University,\\ 
Taipei, Taiwan 10617, ROC}

\emailAdd{phcheng@phys.sinica.edu.tw}
\emailAdd{chengwei@phys.ntu.edu.tw}

\abstract{
Direct \CP violation in two-body $D\to PP$ and $D\to V\!P$ decays is studied within the framework of the topological amplitude approach for tree amplitudes and the QCD factorization approach for penguin amplitudes. It is the interference between tree and {\it long-distance} penguin and penguin-exchange amplitudes that pushes \CP asymmetry difference between $D^0\to K^+K^-$ and $\pi^+\pi^-$ modes up to the per mille level.
Using the same mechanism, we find that \CP asymmetry can also occur at the $10^{-3}$ level in many of the $D\to V\!P$ channels or otherwise be negligibly small. There are six golden modes which have sufficiently large branching fractions and direct \CP violation at the per mille level.  In particular, the direct \CP asymmetry difference between $D^0\to K^+K^{*-}$ and $\pi^+\rho^-$ is predicted to be $(-1.61\pm0.33)\times 10^{-3}$, very similar to the counterpart in the $P\!P$ sector.  The LHCb's observation of \CP asymmetry difference can be explained within the standard model without the need of new physics. The key lies in the long-distance penguin topology (penguin and penguin-exchange) arising
from final-state rescattering.

}

\FullConference{%
  11th International Workshop on the CKM Unitarity Triangle (CKM2021)\\
  22-26 November 2021\\
  The University of Melbourne, Australia
}

\begin{document}
\maketitle

\section{Introduction}
In November of 2011 LHCb announced the first evidence of \CP violation in the charm sector. A nonzero value for the difference between the time-integrated \CP asymmetries of the decays $D^0\to K^+K^-$ and $D^0\to\pi^+\pi^-$ \cite{LHCb:2012}
\be
\Delta A_{CP}\equiv A_{CP}(K^+K^-)-A_{CP}(\pi^+\pi^-)=-(0.82\pm0.21\pm0.11)\%
\en
was reported. The significance of the measured deviation from zero is 3.5$\sigma$. This had triggered a flurry of studies exploring whether this \CP violation in the charm sector was consistent with the standard model (SM) or implies new physics (NP). However, the original evidence of \CP asymmetry difference was gone in 2013 and 2014 when LHCb started to use the muon tag to identify the $D^0$ and found a positive $\Delta A_{C\!P}$ \cite{LHCb:2013}.
In 2019, LHCb finally reported the first observation of
\CP asymmetry in the charm system with the result at the per mille level \cite{LHCb:2019}
\be \label{eq:LHCb:2019}
\Delta A_{CP}\equiv A_{CP}(K^+K^-)-A_{CP}(\pi^+\pi^-)=(-1.54\pm0.29)\times 10^{-3}.
\en

The time-integrated asymmetry can be further decomposed into a direct \CP asymmetry $a_{CP}^{\rm dir}$
and a mixing-induced indirect \CP asymmetry characterized by the parameter $\Delta Y_f$ which measures the asymmetry between $D^0\to f$ and $\bar D^0\to f$ effective decay widths; that is,
$A_{CP}(f,t)\approx a_{CP}^{\rm dir}(f) +(t/\tau(D^0)) \Delta Y_f$.
Based on the LHCb average of $\Delta Y$, it follows that the direct \CP asymmetry difference is given by \cite{LHCb:2019}
\be
\Delta a_{CP}^{\rm dir}=(-1.57\pm0.29)\times 10^{-3}.
\en

In 2012 there existed two independent studies in which direct \CP violation in charmed meson decays was explored based on the topological diagram approach for tree amplitudes and the QCD-inspired approach for penguin amplitudes~\cite{Cheng:2012b,Cheng:2012a,Li:2012}. Interestingly, both works predicted a $\Delta A_{C\!P}$ at the per mille level with a negative sign seven years before the LHCb's announcement of the first observation of \CP violation in the charm sector.

\section{\CP violation in $D\to PP$ decays}
The amplitude of the singly Cabibbo-suppressed charmed meson decay in general reads
\be
A(D\to PP)=\lambda_d({\rm tree+penguin})+\lambda_s({\rm tree'+penguin'}),
\en
where $\lambda_p\equiv V^*_{cp}V_{up}$.
Direct \CP asymmetry can occur at the tree level if the tree amplitudes denoted by ``tree" and ``tree$'$" have a nontrivial strong phase difference. For example, consider the decay $D_s^+\to K^+\eta$ with the quark-diagram amplitudes
\be
A(D_s^+\to K^+\eta)={1\over \sqrt{2}}[\lambda_d(C+P^d)+\lambda_s(A+P^s)]\cos\phi-[\lambda_d P^d+\lambda_s(T+C+A+P^s)]\sin\phi,
\en
where $T,C,A,P$ are color-allowed external $W$-emission, color-suppressed internal $W$-emission, $W$-annihilation and penguin topological amplitudes, respectively, and $\phi$ is the $\eta$--$\eta'$ mixing angle. The tree topological amplitudes can be extracted from Cabibbo-favored $D\to PP$ decays to be (in units of $10^{-6}$ GeV) \cite{Cheng:2019}
\be
T=3.113\pm0.011, && C=(2.767\pm0.029)e^{-(151.3\pm0.3)^\circ},  \non \\
E=(1.48\pm0.04)e^{i(120.9\pm0.4)^\circ}, && A=(0.55\pm0.03)e^{i(23^{+7}_{-10})^\circ},
\en
with $E$ being the $W$-exchange amplitude.
It is obvious that a large DCPV (direct \CP violation) at the tree level can arise from the interference between $\lambda_d C$ and $\lambda_s T$. Our calculation yields \cite{Cheng:2019}
\be
a_{\CP}^{\rm tree}(D_s^+\to K^+\eta)=(-0.75\pm0.01)\times 10^{-3}, \quad a_{\CP}^{\rm total}(D_s^+\to K^+\eta)=(-0.81\pm0.08)\times 10^{-3}.
\en
\CP asymmetry in this mode is dominated by the tree-level one of order $10^{-3}$.
A great merit of the diagrammatic approach is that tree DCPV can be reliably estimated as the magnitude and phase of each tree topology can be extracted from the data.

For $D^0\to \pi^+\pi^-$ and $K^+K^-$, their topological amplitudes read
\be \label{eq:AmpDpipi}
A(D^0\to\pi^+\pi^-) &=& \lambda_d(T+E^d)+\lambda_p(P^p+\PE^p+\PA^p), \non \\
A(D^0\to K^+K^-) &=& \lambda_d(T+E^s)+\lambda_p(P^p+\PE^p+\PA^p),
\en
where summation over $p=d,s$ is understood, $\PE$ and $\PA$ are penguin-exchange and penguin-annihilation amplitudes, respectively, and the superscript $q$ refers to the quark involved in the associated penguin loop.
In these two decays, direct \CP violation arises from the interference between tree and penguin amplitudes.
The complete expression of $\Delta a_{\rm CP}^{\rm dir}$ is given by~\cite{Cheng:2012b}
\be \label{eq:DCPVdiff}
\Delta a_{\CP}^{\rm dir}=-1.30\times 10^{-3} \left( \left| {P^d+P\!E^d+P\!A^d \over T+E^s-\Delta P}\right|_{_{K\!K}}\sin\delta_{_{K\!K}}+\left| {P^s+P\!E^s+P\!A^s\over T+E^d+\Delta P}\right|_{\pi\pi}\sin\delta_{\pi\pi}\right),
\en
where  the parameter $\Delta P$ is defined by $\Delta P \equiv (P^d+\PE^d+\PA^d)-(P^s+\PE^s+\PA^s)$.

For penguin, penguin-exchange and penguin-annihilation amplitudes, we consider two QCD-inspired approaches, namely, perturbative QCD (pQCD) and QCD factorization (QCDF).
The ratio $|P/T|$ is na{\"i}vely expected to be of order $(\alpha_s(\mu_c)/\pi)\sim {\cal O}(0.1)$. It was found that
\be \label{eq:PoverT}
&& \left({P\over T}\right)_{\pi\pi}\approx 0.30 e^{i110^\circ}, \qquad \left({P\over T}\right)_{K\!K}\approx 0.24 e^{i110^\circ},  \qquad\quad {\rm pQCD+FAT},  \non \\
&& \left({P\over T}\right)_{\pi\pi}\approx 0.23 e^{-i150^\circ}, \qquad \left({P\over T}\right)_{K\!K}\approx 0.22 e^{-i150^\circ}, \qquad {\rm QCDF+TDA},
\en
in pQCD in conjunction with the so-called factorization-assisted topological-amplitude approach (FAT) \cite{Li:2012} and in QCDF together the topological diagram approach (TDA) \cite{Cheng:2012b,Cheng:2012a}. In the latter approach, we have taken into account SU(3) breaking effects by having $T_{\pi\pi}=0.96T$, $T_{KK}=1.27T$ and  
\be \label{eq:EdEs}
\mbox{Solution~I:} &&  E^d=1.10\, e^{i15.1^\circ}E ~, \qquad E^s=0.62\, e^{-i19.7^\circ}E
~;   \non \\
\mbox{Solution~II:} &&  E^d=1.10\, e^{i15.1^\circ}E ~, \qquad E^s=1.42\, e^{-i13.5^\circ}E
~, 
\en
where $E^q$ refers to the $W$-exchange amplitude associated with $c\bar u\to q\bar q$ ($q=d,s$) \cite{Cheng:2019}
. Based on light-cone sum rules, the ratios 
\be
\left|{P\over T}\right|_{\pi\pi}=0.093\pm0.011, \qquad \left|{P\over T}\right|_{_{KK}}=0.075\pm0.015, \qquad {\rm LCSR},
\en
were obtained in ~\cite{Khodjamirian:2017}. However, the magnitude of $P/T$ turns out to be of order $(0.22\sim 0.30)$ in the QCD-inspired approaches.

After including $W$-exchange, penguin-exchange and penguin-annihilation contributions, it follows that
\be \label{eq:Ratio,pQCD}
\left({P^s+P\!E^s+P\!A^s \over T+E^d +\Delta P}\right)_{\pi\pi}=0.66\, e^{i134^\circ}, \qquad
\left({P^d+P\!E^d+P\!A^d \over T+E^s-\Delta P }\right)_{_{K\!K}}= 0.45\, e^{i131^\circ},
\en
in pQCD+FAT~\cite{Li:2012}, and
\be \label{eq:Ratio,QCDF}
\left({P^s+P\!E^s+P\!A^s \over T+E^d +\Delta P}\right)_{\pi\pi}=0.32\, e^{i176^\circ}, \quad
\left({P^d+P\!E^d+P\!A^d \over T+E^s-\Delta P }\right)_{_{K\!K}}=
\begin{cases}
0.23\,e^{-i164^\circ} \\
0.23\,e^{i178^\circ}
\end{cases},
\en
in QCDF for Solutions I and II of $W$-exchange amplitudes $E^d$ and $E^s$ \cite{Cheng:2019}.
Comparing Eq. (\ref{eq:Ratio,pQCD}) with Eq. (\ref{eq:PoverT}), we see that the magnitude of the ratio $P/T$ is enhanced by a factor of 2 after including $\PE,\PA$, $E$ and $\Delta P$.
Substituting Eq. (\ref{eq:Ratio,pQCD}) into Eq. (\ref{eq:DCPVdiff}) yields
\be
\Delta a_{\rm CP}^{\rm dir}\approx -1.0\times 10^{-3}, \qquad {\rm pQCD+FAT}.
\en
It is obvious from Eq. (\ref{eq:Ratio,QCDF}) that \CP asymmetry difference is very small, of order $10^{-4}$ or less, in QCDF+TDA owing to the reason that the phases $\delta_{\pi\pi}$ and $\delta_{K\!K}$ are not far from $180^\circ$ in QCDF. Even if the phases are allowed to have $25^\circ$ uncertainties, the resultant $\Delta a_{\rm CP}^{\rm dir}$ is still too small compared to experiment. There is one crucial difference in the treatment of the penguin-exchange amplitude in pQCD and QCDF approaches. In pQCD, the factorizable penguin-exchange amplitude $\PE^f$ is proportional to $\la PP|(\bar uq)_{_{S+P}}|0\ra\la 0|(\bar qc)_{_{S-P}}|D\ra$. It was assumed in~\cite{Li:2012} that $\la P_1P_2|(\bar q_1q_2|0\ra=\la P_1P_2|S\ra\la S|\bar q_1q_2|0\ra$ was dominated by the isosinglet heavier scalar resonancs
such as $f_0(1370), f_0(1500)$ and $f_0(1710)$ with the light resonances $f_0(500), f_0(980)$ and $a_0(980)$ being neglected. 
In QCDF, $\PE^f$ is expressed in terms of the twist-2 LCDA $\Phi_M$ and the twist-3 one $\Phi_m$. This explains why the magnitude of the ratio $P/T$ is enhanced by a factor of 2 after including $\PE,\PA$, $E$ and $\Delta P$ in pQCD+FAT, but no so in QCDF+TDA. 

It is known that in the diagrammatic approach, all the topological tree amplitudes except $T$ are dominated by nonfactorizable long-distance effects. This implies that, contrary to $B$ physics, the underlying mechanism of hadronic charm decays is dominated by nonperturbative physics. This is why even until today we still don't have a QCD-inspired theory to describe the nonleptonic decays of charmed mesons. By the same token, it is natural to expect that penguin topology $P$ and $\PE$ also receive long-distance contributions. Indeed, we have pointed out in \cite{Cheng:2012a} the importance of a resonant-like final-state rescattering which has the same topology as the topological graph $P$ and $\PE$. In 2012 we have made the ansatz that $(P+\PE)^{\rm LD}$ is of the same order of magnitude as $E$ and flavor independent. Very recently, this ansatz was shown to be justified in the diagrammatic approach under the SU(3) limit; that is, $(P+\PE)^{\rm LD}=E^{\rm LD}\approx E$ \cite{Wang:2021}. Then we have
\be \label{eq:PoverT+E}
\left({P^s+P\!E^s+P\!A^s+(P+\PE)^{\rm LD} \over T+E^d +\Delta P}\right)_{\pi\pi} &=& 0.77\, e^{i114^\circ}, \non \\
\left({P^d+P\!E^d+P\!A^d+(P+\PE)^{\rm LD} \over T+E^s-\Delta P }\right)_{_{K\!K}} &=&
\begin{cases}
0.45\,e^{i137^\circ} & \mbox{solution I} \\
0.45\,e^{i120^\circ} & \mbox{solution II}
\end{cases}.
\en
Consequently \cite{Cheng:2019},
\be
\Delta a_{CP}^{\rm dir}=
\begin{cases}
(-1.14\pm0.26)\times 10^{-3} & \mbox{solution I} \cr
(-1.25\pm0.25)\times 10^{-3} & \mbox{solution II}
\end{cases},  \qquad {\rm QCDF+TDA}.
\en

Although both QCD-inspired approaches predicted a $\Delta A_{C\!P}$ of order $10^{-3}$ with a negative sign, the mechanisms responsible for \CP violation are quite different. $\Delta a_{CP}^{\rm dir}$ comes from the interference between the tree $T+E$ and the {\it short-distance} penguin $P+\PE$ amplitudes in pQCD+FAT, while it is the interference between tree and {\it long-distance} penguin amplitudes that pushes \CP asymmetry difference up to the per mille level. As we shall see in Sec. 3 below, these two different mechanisms can be discriminated in $D\to V\!P$ decays.

We thus see that the LHCb's observation of \CP asymmetry difference can be explained within the SM without the need of NP. The key lies in the long-distance penguin topology $P+\PE$ arising
from final-state rescattering. A similar but different idea was proposed in Ref. \cite{Grossman:2019xcj}.
When the amplitudes in Eq. (\ref{eq:AmpDpipi}) were re-expressed in terms of $U$-spin components, $\Delta U=0$ and $\Delta U=1$, it was argued in Ref. \cite{Grossman:2019xcj} that the ratio of $\Delta U=0$ over $\Delta U=1$ matrix elements required a non-perturbative enhancement of the penguin contraction of tree operators in order to accommodate the LHCb measurement of $\Delta a_{CP}^{\rm dir}$.

\begin{table}[t]
\caption{Topological amplitudes obtained from Cabibbo-favored  $D\to VP$ decays.  The amplitude sizes are quoted in units of  $10^{-6}(\epsilon\cdot p_D)$ and the strong phases in units of degrees.
}
\label{tab:CFVPB}
\footnotesize{
\begin{center}
\begin{tabular}{ c c c c c c c c c c}
\hline\hline
$|T_V|$                   &$|T_P|$            &$\delta_{T_P}$          &$|C_V|$                          &$\delta_{C_V}$          &$|C_P|$                  &$\delta_{C_P}$                &$|E_V|$                         &$\delta_{E_V}$     \\
$2.19\pm0.03$ &$3.56\pm0.06$ &$61\pm5$ &$1.69\pm0.04$ &$220\pm3$ &$2.02\pm0.02$ &$201\pm1$    &$0.58\pm0.06$ &$283\pm5$  \\
\hline
$|E_P|$                            &$\delta_{E_P}$   &$|A_V|$                           &$\delta_{A_V}$             &$|A_P|$ &$\delta_{A_P}$ &$\chi^2_{\rm min}$ &fit quality\\
            $1.69\pm0.03$ &$108\pm3$ &$0.23\pm0.02$ &$77\pm5$ &$0.18\pm0.03$  &$111^{+13}_{-10}$           & 7.06 & 6.99\% \\ 
\hline\hline
\end{tabular}
\end{center}
\label{fitting}}
\end{table}

\section{\CP violation in $D\to V\!P$ decays}
For $D\to V\!P$ decays, there exist two different types of topological diagrams since the spectator quark of the charmed meson may end up in the pseudoscalar or vector meson. For topological color-allowed tree amplitude $T$ and color-suppressed amplitude $C$ in $D\to V\!P$ decays, the subscript $P$ ($V$) implies that the pseudoscalar (vector) meson contains the spectator quark of the charmed meson.  For the $W$-exchange amplitude $E$ and $W$-annihilation $A$ with the final state $q_1\bar q_2$, the subscript $P$ ($V$) denotes that the pseudoscalar (vector) meson contains the antiquark $\bar q_2$.

By performing a $\chi^2$ fit to the Cabibbo-favored (CF) $D\to V\!P$ decays, we have extracted the magnitudes and strong phases of the topological amplitudes $T_V,C_V,E_V,A_V$ and $T_P,C_P,E_P,A_P$ from the measured partial widths and found five solutions with local $\chi^2$ minima restricted to $\chi^2_{\rm min} < 10$.~\cite{Cheng:2021}. Although all five solutions generally fit the CF modes well, they led to very different predictions for some of the singly Cabibbo-suppressed (SCS) decays.
Especially, the $D^0\to\pi^0\omega$, $D^+\to \pi^+\rho^0$ and $D^+\to\pi^+\omega$ decays were very useful in discriminating among different solutions. We found that one of the solutions which we called (S3') gives a best accommodation of SCS data, while other solutions were ruled out. The magnitudes and relative phases of various topological amplitudes are listed in Table \ref{tab:CFVPB}. The topological amplitudes of all these solutions respect the hierarchy pattern:
\be \label{eq:tree_hierarchy}
|T_P|>|T_V|\gsim|C_{P}|>|C_{V}|\gsim|E_P|>|E_V|>|A_{P,V}|.
\en

\begin{table}[t]
\caption{Direct \CP asymmetries of singly Cabibbo-suppressed  $D\to V\!P$ decays (in units of $10^{-3}$), where $a_{\rm dir}^{({\rm tree})}$ denotes \CP asymmetry arising from purely tree amplitudes. The superscript (t+p) denotes tree plus QCD-penguin amplitudes, (t+pe+pa+s) for tree plus $\PE,\PA$ and $S$ amplitudes, (t+pe$^{\rm LD}$) for tree plus long-distance $\PE$ amplitude induced from final-state rescattering and ``tot'' for the total amplitude.
The predictions from~\cite{Qin} in the FAT approach with the $\rho-\omega$ mixing are listed in the last column for comparison.
  \label{tab:CPVP}  }
%
\begin{center}
\begin{tabular}{ l c  c c r r c}
\hline\hline
 Mode
   & $a_{\rm dir}^{({\rm tree})}$ &  $a_{\rm dir}^{({\rm t+p})}$ & $a_{\rm dir}^{({\rm t+pe+pa+s})}$ &  $a_{\rm dir}^{({\rm t+pe^{\rm LD}})}$
     & $a_{\rm dir}^{({\rm tot})}$~~~~ & $a_{\rm dir}^{({\rm tot})}$\cite{Qin} \\
\hline
  $D^0\to\pi^+ \rho^-$  & 0 & $-0.00\pm0.00$ &  $-0.011\pm0.000$& $0.77\pm0.22$ & $0.76\pm0.22$ & $-0.03$ \\
  $D^0\to\pi^- \rho^+$  & 0 &  $0.01\pm0.00$ & $0.008\pm0.001$  & $-0.13\pm0.08$ & $-0.11\pm0.08$ & $-0.01$ \\
   $D^0\to\pi^0 \rho^0$  &0 &  $-0.01\pm0.00$ & $-0.004\pm0.000$  & $0.28\pm0.16$ & $0.27\pm0.16$ & $-0.03$ \\
   $D^0\to K^+ K^{*-}$  & 0 &  $-0.01\pm0.01$ & $0.011\pm0.000$ & $-0.85\pm0.24$ & $-0.85\pm0.24$ & $-0.01$  \\
   $D^0\to K^- K^{*+}$   & 0 &  $-0.03\pm0.00$ & $-0.009\pm0.000$  & $0.08\pm0.09$ & $0.04\pm0.09$ & 0  \\
   $D^0\to K^0 \ol{K}^{*0}$   & $-0.03\pm0.02$ &  $-0.03\pm0.02$ & $-0.03\pm0.02$ & $-0.03\pm0.02$ & $-0.03\pm0.02$ & $-0.7$ \\
   $D^0\to\ol{K}^0 K^{*0}$
     & $1.07\pm0.12$ &  $1.07\pm0.12$ & $1.07\pm0.12$ & $1.07\pm0.12$ & $1.07\pm0.12$ & $-0.7$ \\
\hline
   $D^+\to\pi^+ \rho^0$  & 0 &  $0.33\pm0.02$ & $0.10\pm0.01$ & $0.83\pm1.36$ & $1.26\pm1.34$ & 0.5\\
   $D^+\to\pi^0 \rho^+$  & 0 &  $0.10\pm0.01$ & $0.04\pm0.00$ & $-0.58\pm0.52$ & $-0.44\pm0.52$ & 0.2 \\
   $D^+\to\eta \rho^+$ &   $-1.85\pm0.51$ &  $-1.97\pm0.54$ & $-1.93\pm0.55$ & $-2.31\pm0.92$ & $-2.50\pm0.98$ & $-0.6$ \\
   $D^+\to\eta\,' \rho^+$ &  $0.23\pm0.05$ &  $0.20\pm0.05$ & $0.21\pm0.05$ & $0.39\pm0.16$ & $0.34\pm0.15$ & 0.5 \\
   $D^+\to K^+ \ol{K}^{*0}$  &  $-0.11\pm0.01$ &  $-0.14\pm0.01$ & $-0.11\pm0.01$ & $-0.77\pm0.24$ & $-0.80\pm0.24$ & 0.2\\
   $D^+\to\ol{K}^0 K^{*+}$  &  $-0.04\pm0.01$ &  $-0.05\pm0.01$ & $-0.05\pm0.01$ & $-0.06\pm0.06$ & $0.04\pm0.07$ & 0.04 \\
\hline
   $D_s^+\to \pi^+ K^{*0}$ &  $0.18\pm0.02$ &  $0.24\pm0.02$ & $0.19\pm0.02$ & $1.25\pm0.41$ & $1.32\pm0.41$ & $-0.1$  \\
   $D_s^+\to \pi^0 K^{*+}$  &  $0.13\pm0.02$ &   $0.12\pm0.03$ & $0.11\pm0.03$ & $1.35\pm0.40$ & $1.31\pm0.40$ & $-0.2$  \\
   $D_s^+\to K^+ \rho^0$  &  $0.14\pm0.03$ &  $0.11\pm0.02$ & $0.15\pm0.03$ & $-0.26\pm0.12$ & $-0.29\pm0.12$ & 0.3 \\
   $D_s^+\to K^0 \rho^+$  &  $0.06\pm0.02$ &  $0.08\pm0.02$ & $0.08\pm0.02$ & $-0.10\pm0.10$ & $-0.07\pm0.10$ & 0.3 \\
   \hline\hline
\end{tabular}
\end{center}
\end{table}
%

As in the case of $D\to PP$ decays, we make a similar ansatz for the long-distance contributions to $P+\PE_{V,P}$, namely,
\be
(P+\PE_V)^{\rm LD}\approx E_P, \qquad (P+\PE_P)^{\rm LD}\approx E_V.
\en
The calculated \CP asymmetries of the SCS $D\to VP$ decays are exhibited in Table \ref{tab:CPVP}. From Table~\ref{tab:CPVP} we identify several golden modes which have large branching fractions and sizeable \CP asymmetries at the order of $10^{-3}$:
\be \label{eq:golden}
D^0\to \pi^+\rho^-, K^+K^{*-},\qquad D^+\to \eta\rho^+, K^+\overline{K}^{*0},  \qquad
D_s^+\to \pi^+ K^{*0}, \pi^0K^{*+}.
\en
It is interesting to notice that the \CP asymmetry difference defined by
\be \label{eq:acp VP}
\Delta a_{CP}^{V\!P}\equiv a_{CP}(K^+K^{*-})-a_{CP}(\pi^+\rho^-),
\en
in analogy to $\Delta A_{C\!P}$ defined in Eq.~(\ref{eq:LHCb:2019}) for the corresponding $P\!P$ final states, is predicted to be $(-1.61\pm0.33)\times 10^{-3}$, which is very similar to the observed \CP asymmetry difference between $D^0\to K^+K^-$ and $D^0\to\pi^+\pi^-$. This is an attractive and measurable observable in the near future.
It is thus desirable to first search for \CP violation in the aforementioned golden modes.

From Table~\ref{tab:CPVP} it is evident that the predicted \CP asymmetries given in~Ref. \cite{Qin} based on the pQCD+FAT approach are generally smaller than ours by one to two orders of magnitude. In this approach the factorizable penguin-exchange amplitude $\PE^f_{V(P)}$ is proportional to $\la VP(PV)|(\bar uq)_{_{S+P}}|0\ra\la 0|(\bar qc)_{_{S-P}}|D\ra$. It was evaluated in the pole model by assuming its dominance by resonant pseudoscalars. It was shown in~Ref. \cite{Qin} that there was a numerical coincidence that the short-distance $P_V$ and $\PE_V^f$ canceled each other. As a consequence, \CP asymmetries in $D^0\to \pi^\pm\rho^\mp$ and $D^0\to K^\pm K^{*\mp}$ decays are very small of order $10^{-5}$.  They are also very small in QCDF+TDA for a reason quite different from FAT: various phase angles  become smaller or even close to zero after including the contributions from $\PE,\PA$ and $W$-exchange to the ratio of $P/T$.
It is the long-distance penguin topology that explains why our predictions of \CP asymmetries in  $D^0\to \pi^\pm\rho^\mp$ and $D^0\to K^\pm K^{*\mp}$ are much bigger than those in pQCD+FAT.


\end{document}